\documentclass[a4paper, 11pt]{article}
\usepackage{latexsym}
\usepackage{amssymb}
\usepackage{amsmath}
\usepackage{amsfonts}
\usepackage{bbm}
\usepackage{graphicx}
\usepackage{float}
\usepackage[english]{babel}
\usepackage{multirow}
\usepackage{float}
\usepackage{soul,color}
\restylefloat{table}
\usepackage{caption}
\usepackage{placeins}
\usepackage{hyperref}
\usepackage{wrapfig}
\usepackage{cleveref}
\usepackage{enumitem}
\usepackage[toc,page]{appendix}
\usepackage[normalem]{ulem}
\useunder{\uline}{\ul}{}
\usepackage{ltxtable}
\usepackage{subfig}
\usepackage{todonotes}
\setuptodonotes{inline}
\usepackage{subfig}
\usepackage{longtable}
\newcolumntype{L}[1]{>{\raggedright\let\newline\\\arraybackslash\hspace{0pt}}m{#1}}
\newcolumntype{C}[1]{>{\centering\let\newline\\\arraybackslash\hspace{0pt}}m{#1}}

\newcommand{\quota}[1]{``#1''}
\makeatletter
\def\@maketitle{
\begin{center}
{\Huge \bfseries \sffamily \@title }\\[3ex]
{\Large  \@author}\\[2ex]
\includegraphics[width = 30mm]{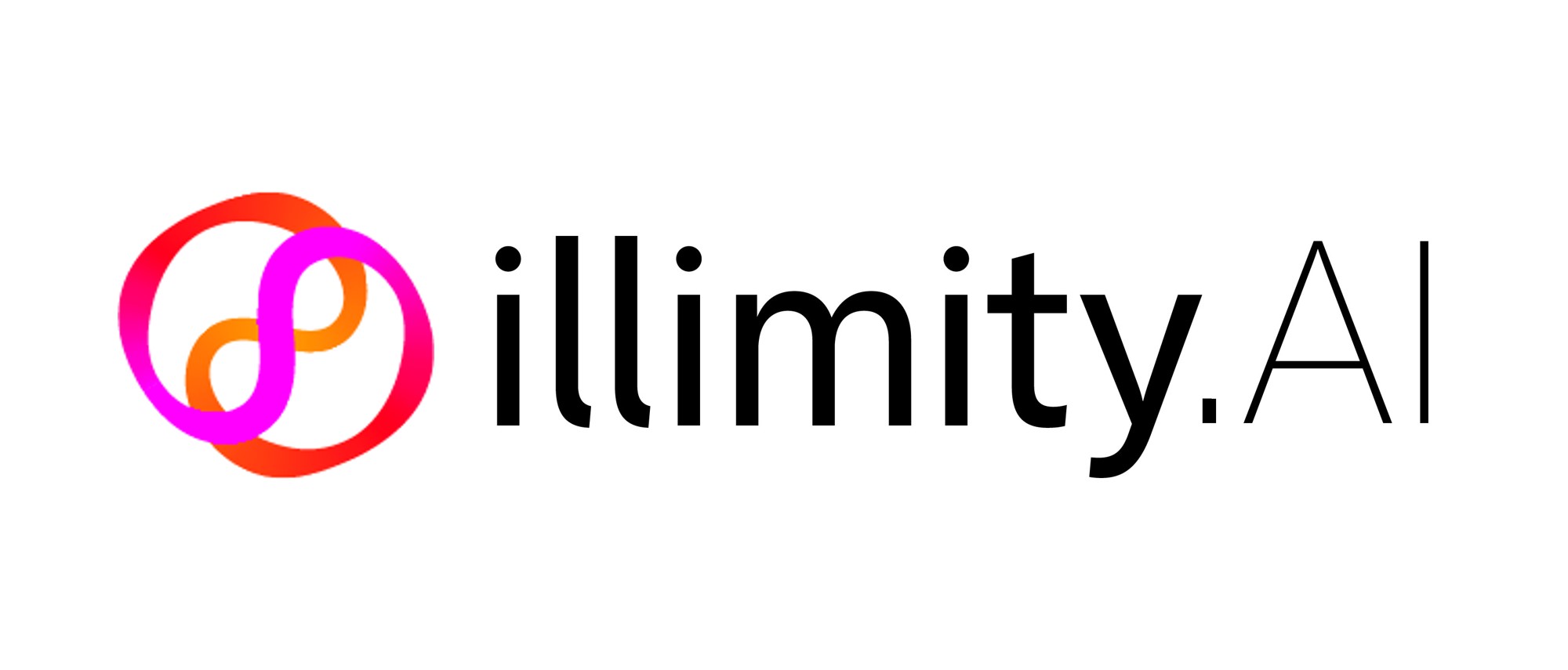}\\[4ex]
\@date\\[4ex]
\end{center}}
\makeatother
\author{O. Didkovskyi\thanks{corresponding author: oleksandr.didkovskyi@illimity.com}, N. Jean, G. Le Pera and C. Nordio}
\title{\bf Cross-Domain Behavioral Credit Modeling: transferability from private to central data}

\begin{document}\maketitle

\vspace{10mm}

\begin{abstract}
\noindent

This paper introduces a credit risk rating model for credit risk assessment in quantitative finance, aiming to categorize borrowers based on their behavioral data. The model is trained on data from Experian, a widely recognized credit bureau, to effectively identify instances of loan defaults among bank customers. Employing state-of-the-art statistical and machine learning techniques ensures the model's predictive accuracy. Furthermore, we assess the model's transferability by testing it on behavioral data from the Bank of Italy, demonstrating its potential applicability across diverse datasets during prediction. This study highlights the benefits of incorporating external behavioral data to improve credit risk assessment in financial institutions.

\end{abstract}

\bigskip
{\bf JEL} Classification codes: C45, C55, G24
{\bf AMS} Classification codes: 91G40
\bigskip


\newpage
\tableofcontents
\newpage

%%%%%%%%%%%%%%%%%%%%%%%%%%%%%%%%%%%%%%%%%%%%%%%%%%%%%%%%%%%%%%%%%%%%%%%%%%%%%%%%%%%%%%%%%

\section*{Introduction}
\label{section:intro}
\addcontentsline{toc}{section}{Introduction}

Corporate credit risk assessment often relies on financial statements information and behavioral data.
However, financial statements are typically updated on a yearly basis, which means they do not capture within-year variability of financial and economic indicators.
Moreover, there is usually a delay in obtaining the balance sheet for a given fiscal year, as it becomes available only after board approval, which can be several months later.

Behavioral models, on the other hand, estimate the creditworthiness of an obligor by leveraging information
from banking credit lines, including bank account balances, overdrafts, credit card limits, and loan installment
delays. For large corporations, credit ratings are primarily driven by financial statements and annual reports,
while for small to medium-sized corporations, events and trends related to bank loan payments, past due amounts,
and delinquency status play a more significant role.
Regulatory guidelines worldwide suggest considering both these components in corporate credit risk assessment.
Banks often develop their internal rating models by training them on their internal time series of behavioral data, along with external data and scorecards
provided by national central banks and credit reporting companies that collect data from multiple sources,
including the banks themselves.

In recent years, usage of machine learning (ML) and deep learning (DL) models, combined with information
from data providers, has gained popularity due to their high performance and ability to uncover hidden
non-linear dependencies \cite{khandani2010consumer, butaru2016risk, sirignano2018deep, albanesi2019predicting, provenzano2020machine, provenzano2019artificial}. For instance, \cite{albanesi2019predicting}
proposes to use Experian data for consumer credit risk assessment, estimating the probability of a past due 90 in the next two years and establishing a
baseline performance for consumer risk analysis.

In this paper, we build a credit scoring behavioral model for corporate clients, using external Experian data sets as training data.
The goal is to provide an accurate estimate of borrowers creditworthiness, similar to the approach presented in \cite{provenzano2020machine},
with the main differences being the data used and the target variable.
Our solution consists of three consecutive models: the first predicts the probability of default, the second calibrates the estimated probabilities,
and the last one builds a rating master scale.
This ensures that the solution can be effectively used for credit ranking, pricing, and corporate lending.
Furthermore, we introduce a potential feature mapping technique to enable the utilization of the calibrated model with publicly accessible data from the Central Credit Register. This registry comprehensively captures data from the entire national banking system, offering a complete overview of each client's credit exposure by design.
To ensure the correctness of the mapping we set up a rigorous dual-strategy approach.
Firstly, we affirm existence of the expected connection between these two data sources via properly defined multiple hypothesis testing with the False Discovery Rate control \cite{benjamini1995controlling, benjamini2001control}.
Then, we verify the effectiveness of the model on the the Central Credit Register data through carefully managed backtesting procedure.

This article starts with an analysis of the dataset in section \ref{section:data}, where we provide a comprehensive description,
focusing on the Experian data used in the study and highlighting its main categories.
Following this, in section \ref{section:features} we establish feature engineering and feature selection procedures.
Section \ref{section:baseline_approaches} introduces baseline approaches for validating the effectiveness of the proposed
behavioral rating model.
Section \ref{section:model} then delves into the ML architecture we employed.
The results, including model metrics and explainability using Shapley values, are presented in section \ref{section:results}.
Finally, in the last section, we demonstrate the application of our model to the Central Credit Register data with subsequent validation. %of the transfer learning.

\section{Experian dataset description}
\label{section:data}

The Experian database is updated on a monthly basis to track all recent changes and fluctuations.
We utilize historical data for two lengthy periods: June 2017 to June 2019 and March 2020 to March 2021,
with monthly updates. The database contains information on approximately 1.7 million Italian legal entities,
of which around 60\% are from the corporate sector and the remaining 40\% are private entities.
Each monthly snapshot consists of 266 columns.% (see Appendix \ref{appendix:appendix_experian_variables_description})

Experian data can be grouped into five categories:

\begin{itemize}
     \item Credit Account Previous Searches (CAPS): this subdataset contains information about credit requests made during the past several months.
     \item Credit Account Information Sharing (CAIS): this subdataset provides a snapshot of a company's current financial state.
     It includes various numeric characteristics, such as the number of active credit lines and the balance of mortgages and installments.
     These variables can be classified into two main classes of credit lines: open credits (referred to as NRT) and installments/mortgages (referred to as RT).
     The data within this subdataset are essential for our model's features.
     Additionally, this subdataset provides indicators of the worst status among all contracts, indicating the
     presence of non-performing loans (NPL) or insolvency.
     \item Protests and Prejudicial Claims: this category represents information about existing legal processes related to a company.
     \item Public Data: this category contains additional public data related to a company.
     \item Experian indicators and scores: this category includes different credit scores such as the DG3r and the DCM \cite{experian}.
\end{itemize}

The dataset includes both raw data and pre-processed features that Experian deems important for predicting credit events. It is indeed the dataset used at prediction time by Experian models and not the raw dataset used for training their model.
Some features capture trends over time, while others show relationships between different features, such as the draw ratio percentage for NRT products. The large number of features is due to the inclusion of these pre-processed values.

\subsection{Target definition}
\label{subsection:target}
Our binary target is based on the regulatory default definition, which identifies any credit state that is
equal to or worse than 90 days past due in the last 12 months. It encompasses the following borrower credit status \cite{bankit_web}:
\begin{itemize}
     \item Bad Loans: exposures to debtors that are insolvent or in substantially similar circumstances.
     \item Unlikely-to-pay (UTP) exposures: these are aside from bad loans and include cases where banks believe the debtors are unlikely
     to meet their contractual obligations in full unless certain actions, such as enforcing guarantees, are taken.
     Experian's closest definition is when at least one contract is in default, which implies a change of status to a non-performing loan (NPL) with subsequent closure of the contract.
     \item Past due: aside from those classified as bad loans and unlikely-to-pay exposures, this category includes overdrafts and/or past-due amounts
     of more than 90 days and above a predefined threshold.
     Experian allows us to identify two types of contracts in this state: installment contracts (RT) and non-installment contracts (NRT).
\end{itemize}
These definitions form the basis of our target variable for the binary classification of default.

\subsection{Dataset Filtering}
\label{subsection:dataset_filtering}
An extensive dataset was created by collecting data from five different time periods: March 2018, 2019, 2020, 2021, and 2022. Our analysis specifically focused on active companies that did not have any records of \quota{insolvency} as their worst special status, ensuring a more stable dataset. Thus, we excluded the following categories from each dataset:

\begin{itemize}
\item Data related to private individuals, more suited to build a retail model than a corporate one.
\item Companies with no active contracts (CAIS), having therefore no contract on which they can default.
\item Companies that were insolvent 1 year prior (in the previous dataset)
\end{itemize}

\begin{figure}[h!]
     \centering
	\includegraphics[scale=0.6]{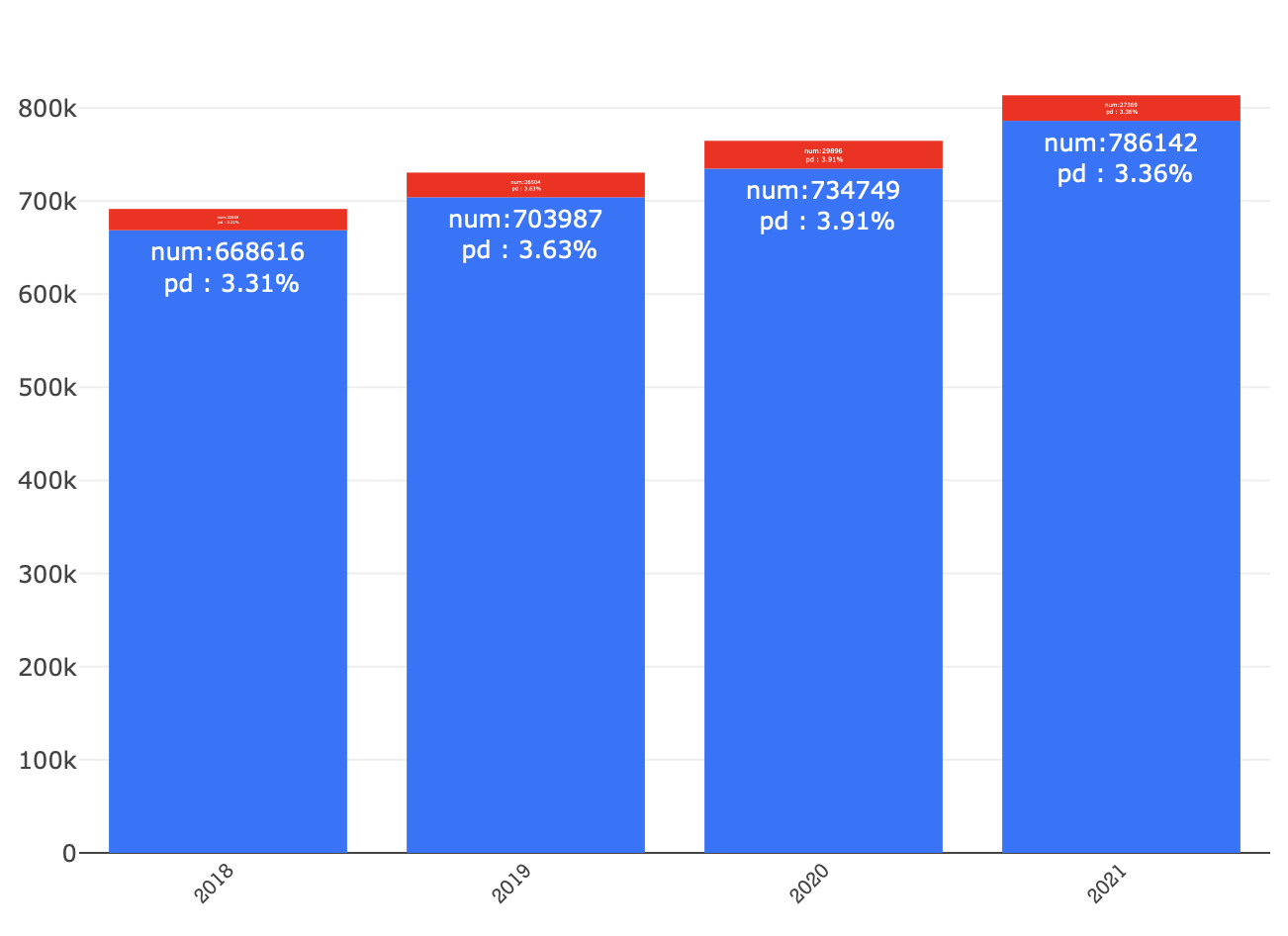}
     \caption{Number of companies and default rate by year}
     \label{fig:observations_by_year}
\end{figure}

The final training dataset, post-filtering, comprised 2.9 million records. The average default rate of 3.5\% suggests a considerable imbalance, coherent with the historical default rate for italian companies published by the Bank of Italy \href{https://infostat.bancaditalia.it/inquiry/home?spyglass/taxo:CUBESET=&ITEMSELEZ=&OPEN=true/&ep:LC=IT&COMM=BANKITALIA&ENV=LIVE&CTX=DIFF&IDX=1&/view:CUBEIDS=TRI30605/}(infostat). Figure \ref{fig:observations_by_year} shows the number of observations and the default (target) rates for each year. A longer time span cannot be achieved due to Experian data retention policy that implies a maximum retention period of historical data up to 36 months.

Upon data filtering, we implemented out-of-sample and out-of-time validation, splitting the dataset twice:
80\%-20\% for (stratified) out-of-sample, and datasets from 2018-2021 versus 2022 for out-of-time.

\FloatBarrier

\section{Feature engineering and feature selection}
\label{section:features}

As stated above, the Experian dataset comprises 266 columns, related to loans and payments spanning multiple times and multiple product types.
Before engaging in a training of a model we need to cleanse the dataset and harmonize the available features to the ones we get at prediction time, both from Experian and the Central Credit Register, since our model needs to be able to run on both data sources.

The initial and most straightforward data cleaning step involves eliminating features that contain more than 20\% missing values (NaNs).
We adopt this approach because, during the prediction phase, there is a considerable probability of encountering missing values for these features. While these features may serve a valuable purpose during the model training phase, they could potentially mislead the model approximately one fifth of the time when applied to real-world data.

We further enhance data quality by removing collinear features identified through the Variance Inflation Factor (VIF) method  \cite{draper1998applied}.
The presence of such features can introduce noise during model training and offers no discernible benefits.
During this process, a degree of human judgment is exercised to prioritize retaining more interpretable features when the algorithm exhibits uncertainty in selecting the feature to be dropped.

As final step of the algorithmic selection we apply a modified version of the Boruta algorithm, which utilizes LightGBM and computes feature importance using Shapley values, to identify redundant or uninformative variables.
Consequently, the size of the feature set has been reduced from 55 to just 20, thereby simplifying the model and enhancing its interpretability.

Categorical features undergo a distinct treatment in our approach.
We employ James-Stein target encoding for most categorical variables, with one notable exception being the 'NACE' category.
For 'NACE' we adopt a method outlined in \cite{provenzano2020machine}, wherein NACE descriptions are transformed into 5-dimensional vectors using sentence embedding and a stacked autoencoder.

This unique NACE embedding technique offers several advantages.
Firstly, it yields a semantically meaningful representation, preserving sector similarities that may be lost when employing JamesStein encoding.
Additionally, it provides a flexible embedding based on the description rather than static categories.
This flexibility proves invaluable during prediction, especially when dealing with companies spanning multiple sectors, where a single NACE code may not adequately capture the firm's market complexity.

\subsection{Feature generation}

To enable the application of a model trained on Experian data to the Central Credit Register database, a critical step involves aligning these two datasets.
Experian's dataset relies heavily on specific factors such as loan maturity, payment frequency, collateral type, and the number of contracts.
In contrast, the Central Credit Register data is categorized more broadly based on risk categories like self-liquidating risks, maturity risks, and revocable credit risks (\quota{rischi autoliquidanti}, \quota{rischi a scadenza}, and \quota{rischi a revoca} in Italian).

Experian's dataset contains more granular information compared to the Central Credit Register.
Therefore, by appropriately grouping and transforming the Experian data, we aim to create a mapping that closely aligns it with the Central Credit Register's dataset.
Our objective is to achieve this alignment without significantly compromising the model's predictive performance for loan default risk assessment.

These are the main transformations:
\begin{itemize}
\item $RT\_balance$ = $RT\_mortgages\_balance$ + $RT\_non\_mortgages\_balance$
\item \quota{Past\_due\_0\_contracts\_12\_months} to a binary flag
\item \quota{Number of contracts in default} ($def\_no$) to a binary flag
\item \quota{Max\_past\_due\_days\_6\_months} in buckets 0, 1, 30, 90, 180, and more than 180
\end{itemize}

The other Experian features related to the same information listed above but on different years are dropped, reducing drastically the number of features.
Some Key Performance Indicators (KPIs) are then added to help the model training and create features easier to understand by credit analyst:
\begin{itemize}
\item \quota{NRT RT ratio} - the ratio of NRT balance to RT balance
\item \quota{NRTused RT ratio} - the ratio of used NRT balance to RT balance
\item \quota{Draw Ratio NRT} - the ratio of used NRT balance to NRT balance
\end{itemize}

\subsection{Selected Features}

\begin{figure}[h!]
     \centering

	\includegraphics[scale=0.4]{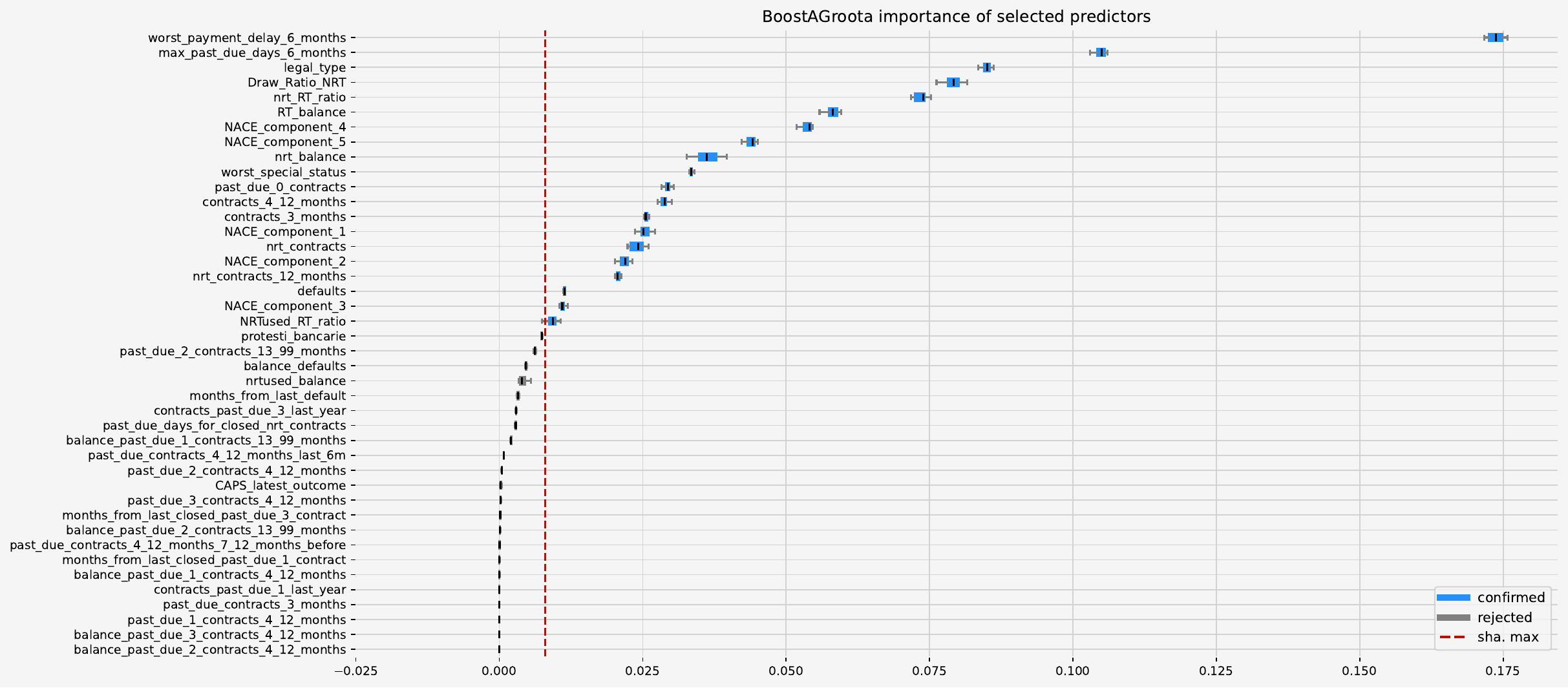}
     \caption{Feature selection importance, grey dashed line represents shap importance threshold }
     \label{fig:feature_importance}
\end{figure}

As illustrated in Figure \ref{fig:feature_importance}, the most influential feature is \quota{RT past due 6} or \quota{Worst\_payment\_delay\_6}, a numeric feature that indicate number unpaid installments. This aligns with intuition, as past payment behavior is often a strong predictor of future credit risk.

The second most important feature, \quota{Max\_past\_due\_days\_6\_months}, is an indicator of the maximum number of day of past due for open credits. Once again, it is logical to assume that past difficulties in maintaining timely payments could predict future payment delinquencies.

Additional notable features include \quota{Draw\_Ratio\_NRT} and \quota{NRT\_RT\_ratio}. These variables reflect the usage of open credits and their proportion in comparison to all credit lines, respectively. It is noteworthy that these findings align well with the predictions of credit analysts, providing further support for the effectiveness of the BoostAGroota approach for feature selection.

\FloatBarrier

\subsection{Final set of features}
\label{subsubsection:final_set}
The feature selection process results in rigorously selected set of meaningful features that are coherent with the proposed target.
A high portion of them are binary indicators that characterize company's state from various perspectives, whereas other variables represent RT, NRT balances and the associated KPIs.
\begin{itemize}
     \item \quota{Special status} - Worst special status, (e.g. insolvency, NPL) over all contracts
     \item \quota{NPL present} ($def\_no>= 1$)   - a flag that indicates that at least one contract became NPL during last 12 months
     \item \quota{Closed past due 0} (\quota{Past\_due\_0\_contracts\_12\_months} $>= 1$) - a flag that indicates that at least one contract without overdue was closed during last 12 months
     \item \quota{Closed NRT} (\quota{NRT\_contracts\_12\_months} $>= 1$) - a flag that indicates that at least one  NRT contract was closed during last 12 months
     \item \quota{Contracts 3 months} - a flag that indicates that at least one contract was opened during last 3 months and it is still active
     \item \quota{Contracts 4 12 months} - a flag that indicates that at least one contract was opened during last 4-12 months and it is still active
     \item \quota{RT past due} (\quota{Worst\_Payment\_Delay\_6\_months}) - maximum number of unpaid installments
     \item \quota{Past due 0} (\quota{Past\_due\_0\_contracts}$>= 1$) - a flag that indicates that at least one contract is in bonis at the moment
     \item \quota{NRT present} (\quota{NRT\_contracts} $>= 1$) -  a flag that indicates that at least one NRT contract is active
     \item \quota{NRT balance} - total balance of active NRT contracts
     \item \quota{NRT past due} (bins of \quota{Max\_past\_due\_days\_6\_months}) - maximum number of past due days for NRT contracts during last 6 months, binned in the following categories: 0, 5, 30, 60, 90, 120, 180
     \item \quota{Protesti} (\quota{protest\_present}) - a flag that indicates presence of at least one cheque protest
     \item \quota{Legal type} - company legal type that can be sole proprietorship (dittà individuale in Italian, hereafter DI), limited company (società di capitali in Italian, hereafter SC), or a partnership (società di persone in Italian, hereafter SP)
     \item \quota{RT balance} - total balance of active mortgages and installments
     \item \quota{NRT RT ratio} - ratio between NRT balance and RT balance
     \item \quota{NRTused RT ratio} - ratio between used NRT balance and RT balance
     \item \quota{Draw Ratio NRT} - ratio between used NRT balance and NRT balance
     \item \quota{Nace embeddings} - the nace embeddings are forced to be selected as a group
\end{itemize}

\section{Baseline approaches for binary classification}
\label{section:baseline_approaches}

In this section, we present three approaches to establish a baseline for our model.
The first is the DG3r index provided by Experian, it is considered as a practical industry benchmark.
The second, a logistic regression with step-wise forward Selection, represents a traditional statistical approach.
The third approach, based on the use of AutoGluon \cite{erickson2020autogluon}, shows the potential of Automated Machine Learning (AutoML).
These approaches were chosen to provide a comprehensive analysis between standard market practices, statistical approaches, and advanced machine learning techniques.
Table \ref{tab:baseline_metrics} reports the metrics of the baseline approaches.

\subsection{Experian Score}
Experian currently generates a credit score known as DG3r, designed to estimate the probability of default within 1 year.
This score varies between 0 and 600 and is associated with a 20 bins rating scale.
However, despite its simplicity and clarity, the index faces limitations, including validation performed on Experian internal dataset and a lack of explainability.

The DG3r score can be transformed into a probability of default by applying logistic transformation with specific parameters. With these values and our target, we can assess DG3r's effectiveness using the same machine learning metrics employed with our model (see Section \ref{section:results}). For instance, DG3r's AUC values for each of the five datasets are approximately 0.86.

Using the DG3r model alone presents challenges in achieving both explainability and suitability for the Central Credit Register data. Consequently, this model is primarily employed to establish a performance baseline.

\subsection{Logistic Regression with Stepwise-Forward Selection}

In this case, we opted to include almost all available variables and conducted automatic feature processing and selection. However, the primary limitation of this approach is the absence of nonlinear interactions within the model, leading to an AUC of 0.85

\subsection{AutoML: AutoGluon}

As expected, the AutoML model exhibited approximately 5\% better performance in terms of the main metrics discussed in Section \ref{section:results}, reaching an AUC of 90\%. However, the final solution is excessively complex, comprising three layers of stacking and over 20 models with various interactions.

The explainability of the AutoGluon model is intricate and questionable for our application, considering the industry's demand for clarity, accuracy, and rigor.

\section{Solution architecture}
\label{section:model}

To develop a reliable solution for effective corporate risk assessment and evaluation with optimal performance, explainability, and transferability to the Central Credit Register data, we have formulated a three-step approach, building upon the methodology presented in \cite{provenzano2020machine}. Each step has undergone significant improvements, incorporating various methodological developments.

The first step employs Gradient Boosting Decision Trees \cite{hastie2009elements} to create a binary classification model. The objective is to identify companies with a high risk of being past due for 90 days or classified as Unlikely to Pay (UTP). This model allows for the identification of companies that may pose a higher credit risk.

In the subsequent stage, the predicted probability of default obtained from the binary classification model undergoes a calibration process. This calibration aligns the scale of the predicted PDs with the actual scale of default rates.

Finally, the calibrated PDs are clustered to establish robust binning and rating classes. This clustering process enables the creation of distinct risk categories for credit risk assessment. These rating classes provide a comprehensive framework for evaluating the creditworthiness of companies.

Through this three-step approach, we have enhanced the methodology for corporate risk assessment, delivering a robust and accurate solution suitable for effective deployment in production environments.

\subsection[Step 1: Past due binary classification]{Past due binary classification}

Similarly to \cite{provenzano2020machine}, we emphasize the importance of a tailored evaluation metric, specifically the $F_{\beta}$ score.
This metric was adopted because it balances specificity and recall with additional weight given to the former, hence reflecting the equilibrium between minimizing credit risk and maximizing financial returns.
\begin{equation}
     F_{\beta} = (1+\beta^2)\frac{\text{specificity} \cdot \text{recall}}{\beta^2 \cdot \text{specificity} + \text{recall}}
\end{equation}

We have opted for LightGBM \cite{ke2017lightgbm}, a gradient boosting framework, as our classification algorithm,
based on its successful application in our previous projects \cite{provenzano2020machine} and its proven efficiency in handling large datasets.
This choice is reinforced by LightGBM's ability to effectively manage imbalanced data.

The performance of the model is contingent on the selected set of hyperparameters. To identify an optimal set, we leverage Optuna \cite{akiba2019optuna}, a hyperparameter optimization framework based on Bayesian techniques. To ensure the robustness and trustworthiness of the obtained hyperparameters, the optimization is conducted within a time series cross-validation framework \cite{bergmeir2018note}.

\subsection[Step 2: Probability calibration]{Probability calibration}

A classifier trained on an unbalanced dataset typically accurately predicts the order for multiple observations.
However, individual probability of default values might be imprecise and deviate significantly from the actual default rates. This challenge can be mitigated by employing a statistical procedure known as probability calibration \cite{Caruana2012}. It is essential to note that the ordering of the PDs remains unaffected by the probability calibration, and as a result, the AUC does not change post-calibration.

In this context, we utilize beta-calibration \cite{kull2017beta} as one of the calibration approaches. Beta-calibration proves effective in optimizing proper scoring rules \cite{scoring2007}, such as the Brier score:

\begin{equation}\label{eqn:brier}
     \text{BS} = \frac{1}{\text{N}}\sum_{t=1}^N (f_t - o_t)^2
     \end{equation}

\subsection[Step 3: Rating attribution via genetic algorithm]{Rating attribution via genetic algorithm}

Completing the process, PD grouping is executed for subsequent rating attribution.
The binning procedure follows \cite{provenzano2020machine}, incorporates specific adjustments to improve the optimization algorithm's convergence efficiency.

\section{Results}
\label{section:results}
\subsection{Default Classification}

To evaluate the model's ability to classify defaults accurately, we employed a range of established metrics, including AUC (Area Under the Curve), Recall, and Average Precision.
These metrics, discussed in detail in \cite{hastie2009elements, provenzano2020machine}, are crucial for a comprehensive understanding of the model's ability to differentiate between default and non-default classes.

\begin{figure}[h!]
     \centering
	\includegraphics[scale=0.7]{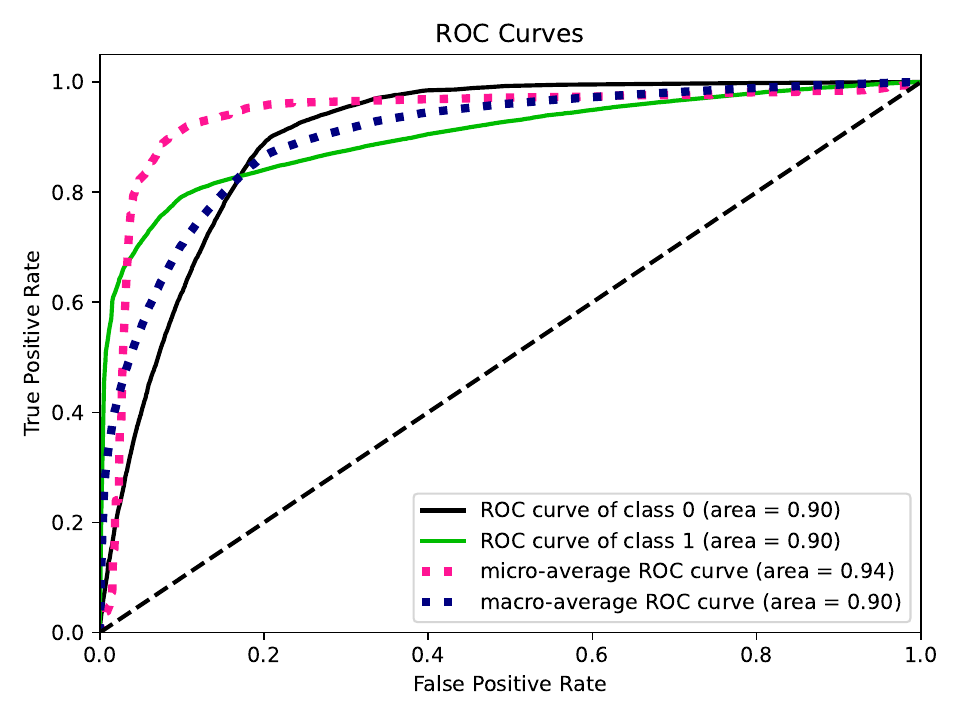}
     \caption{ROC curve for Light-GBM classifier}
     \label{fig:class_ROC}
\end{figure}

The results, detailed in Table \ref{tab:metrics}, highlight the model's performance, showcasing its strong capability in accurately classifying defaults.
Specifically, it achieves an AUC of approximately 0.92 in both out-of-sample and out-of-time validation, as depicted in Figure \ref{fig:class_ROC}.
This high AUC score underlines the model's robust ranking capability.

\begin{table}[h!]
     \centering
        \subfloat[Out-of-sample]{
             \begin{tabular}{|c|c|}
             \hline
                  \textbf{Description} & \textbf{Value} \\ \hline
                  AUC & 0.927 \\ \hline
                  Recall & 0.906 \\ \hline
                  Specificity & 0.717 \\ \hline
                  $F_{\beta}$ Measure & 0.813 \\ \hline
                  F1 Value & 0.205 \\ \hline
                  Average Precision & 0.684 \\ \hline
             \end{tabular}
        }
        \subfloat[Out-of-time]{
             \begin{tabular}{|c|c|}
             \hline
                  \textbf{Description} & \textbf{Value} \\ \hline
                  AUC & 0.920 \\ \hline
                  Recall & 0.907 \\ \hline
                  Specificity & 0.674 \\ \hline
                  $F_{\beta}$ Measure & 0.790 \\ \hline
                  F1 Value & 0.174 \\ \hline
                  Average Precision & 0.664 \\ \hline
             \end{tabular}
        }
   \caption{PD classification model metrics}
   \label{tab:metrics}
   \end{table}
   
\FloatBarrier

Table \ref{tab:baseline_metrics} reports comparison with the baseline models, highlighting its superiority in terms of all used metrics.

\begin{table}[h!]
     \centering
     \begin{tabular}{lcccccc}

     & \textbf{AUC} & \textbf{Recall@0.5} & \textbf{$F_{\beta}$ score} & \textbf{Average Precision} \\
     \textbf{Experian DG3r} & 0.86 & 0.15 & 0.24 & 0.4 \\
     \textbf{AutoGluon} & 0.90 & 0.49 & 0.63 & - \\
     \textbf{Logistic Regression} & 0.85 & 0.42 & 0.56 & 0.56\\
     \textbf{Our Model} & 0.92 & 0.91 & 0.79  & 0.66 \\
     \end{tabular}
     \caption{Comparison with the baseline on our target}
     \label{tab:baseline_metrics}
\end{table}

For further insights into the classification results, we recommend consulting the confusion matrix in table \ref{tab:class_confusion_matrix} from the out-of-time validation.

\begin{table}[h!]
     \centering
     \begin{tabular}{lcc}
          & Predicted Negative (in bonis) & Predicted Positive (defaulted)   \\
          Actual Negative (in bonis) &  0.67 & 0.33  \\
          Actual Positive (defaulted)  & 0.09  & 0.91  \\
     \end{tabular}
     \caption{Normalized confusion matrix with $50\%$ threshold for Light-GBM classifier}
     \label{tab:class_confusion_matrix}
\end{table}

\subsection{PD calibration results}

\begin{figure}[h!]
     \centering
     \subfloat{\includegraphics[scale=0.4]{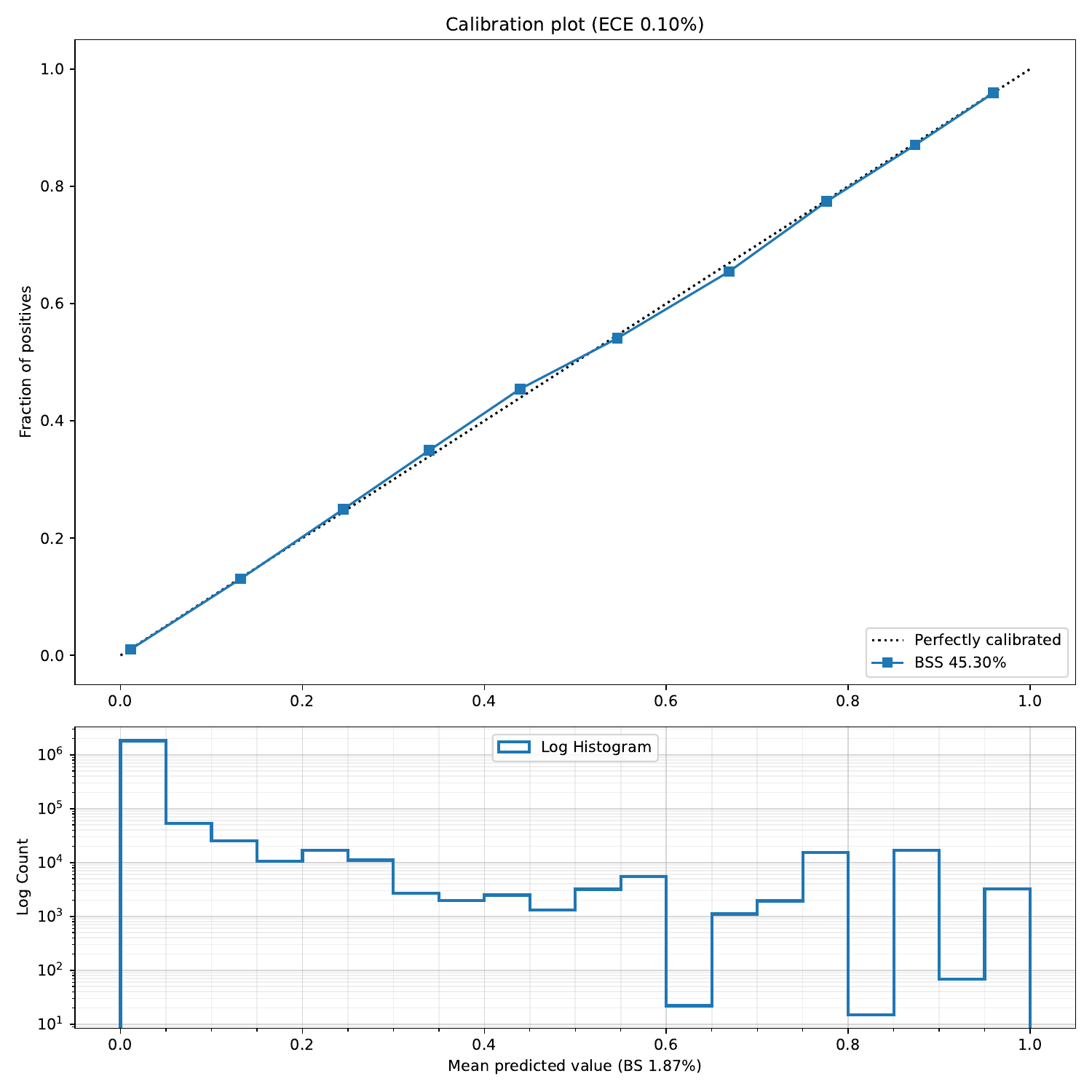}\label{fig:reliability_curve}}
     \caption{Calibration plots and log-scaled histograms of forecast probability
     after (\Cref{fig:reliability_curve}) refitting. Accuracy of predicted probabilities is expressed in term of \emph{log loss} measure}
     \label{fig:calibration_plot}
\end{figure}

\begin{figure}[h!]
     \centering
	\includegraphics[scale=0.7]{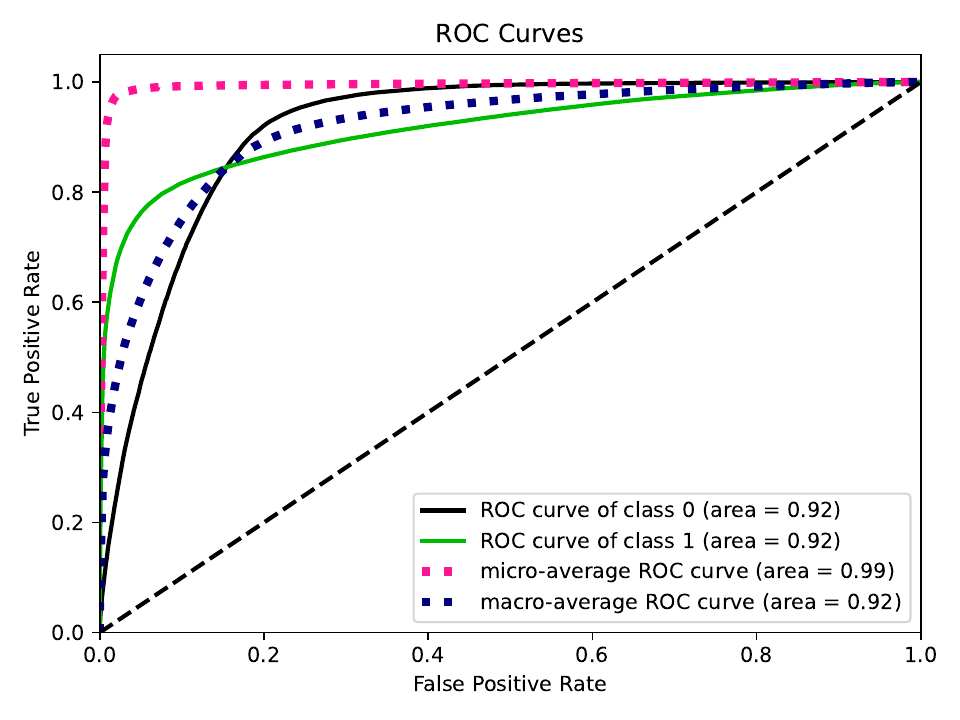}
     \caption{ROC curve for the calibrated classifier}
     \label{fig:pd_ROC}
\end{figure}

Figure \ref{fig:calibration_plot} displays the results of calibration. The reported Brier score is 1.9\%, while the Brier Skill score is 49.3. These figures indicate that the predicted probabilities are notably comparable to the actual ones, as depicted in Figure \ref{fig:pd_ROC}.

Figure \ref{fig:calibration_plot} presents the calibration results of our model, essential for understanding how well the predicted probabilities align with actual outcomes.
The model achieves a Brier score of 1.9\%; a lower Brier score indicates more accurate predictions.
Additionally, the Brier Skill score, standing at 49.3\%, further reinforces the model's reliability.
These metrics suggest that the predicted probabilities generated by our model align closely with the observed outcomes, as further illustrated in Figure \ref{fig:pd_ROC}.

\subsection{Rating classes}
Adopting the methodology outlined in \cite{provenzano2020machine}, we employ a metaheuristic evolutionary algorithm, specifically Differential Evolution, for clustering probability of default values.
This process entails solving a constrained optimization problem using the aforementioned algorithm.
Consistent with practices in credit risk modeling, to validate the resulting rating classes, we apply the Extended Traffic Light Approach \cite{tasche2003traffic} and conduct a Binomial Test.

\begin{equation}\label{eq:traffic_light}
     \begin{cases}
     \mbox{Green} & p_{k} < PD_{k}\\
     \mbox{Yellow} & PD_{k} \leqslant p_{k} < PD_{k}+ K^{y}\sigma(PD_{k}, N_{k})\\
     \mbox{Orange} & PD_{k}+ K^{y}\sigma(PD_{k}, N_{k}) \leqslant p_{k} < PD_{k}+ K^{0}\sigma(PD_{k}, N_{k})\\
     \mbox{Red} & PD_{k}+ K^{0}\sigma(PD_{k}, N_{k}) \leqslant p_{k}
     \end{cases}
     \end{equation}

where $\sigma(PD_{k}, N_{k}) = \sqrt{\frac{PD_{k}(1-PD_{k})}{N_{k}}}$. The parameters $K^{y}$ and $K^{0}$ play a major role in the validation assessment, so they have to be tuned carefully.
A proper choice based on practical considerations is setting $K^{y}= 0.84$ and $K^{0}= 1.44$, which corresponds to a probability of observing green of $0.5$, observing yellow with $0.3$, orange with $0.15$, and red with $0.05$.

Table \ref{tab:rating} reports the obtained rating classes, classes' PDs and out-of-time default rates. 
Noticeably, the Extended Traffic Light Approach results Red for the classes \textit{AA} and \textit{C}, that are edge cases with low numerosity.

\begin{table}[h!]
     \begin{tabular}{c|c|c|c|c|c}
          \hline
          \textbf{\begin{tabular}[c]{@{}c@{}}Rating\\class\end{tabular}} &
          \textbf{\begin{tabular}[c]{@{}c@{}}PD Bins\\(\%)\end{tabular}} &
          \textbf{\begin{tabular}[c]{@{}c@{}}Rating Class\\PD (\%)\end{tabular}} &
          \textbf{\begin{tabular}[c]{@{}c@{}}Out-of-time\\Default Rate\\(\%)\end{tabular}} &
          \textbf{\begin{tabular}[c]{@{}c@{}}One-sided\\Binomial Test\end{tabular}} &
          \textbf{\begin{tabular}[c]{@{}c@{}}Extended\\Traffic Light\\Approach\end{tabular}} \\ \hline
          AAA & {[}0.000, 0.001)  & 0.001  & 0.001  & Passed & Yellow \\
          AA  & {[}0.001, 0.003)  & 0.003  & 0.002  & - & Red \\
          A   & {[}0.003, 0.006)  & 0.004  & 0.005  & Passed & Green \\
          BBB & {[}0.006, 0.020)  & 0.009  & 0.009  & Passed & Green \\
          BB  & {[}0.020, 0.029)  & 0.024  & 0.023  & Passed & Green \\
          B   & {[}0.029, 0.119)  & 0.053  & 0.046  & Passed & Green \\
          CCC & {[}0.119, 0.148)  & 0.133  & 0.132  & Passed & Green \\
          CC  & {[}0.147, 0.564)  & 0.318  & 0.320  & Passed & Green \\
          C   & {[}0.564, 1.000)  & 0.818  & 0.878 & - & Red \\ \hline
     \end{tabular}
     \caption{Out-of-time results of PD clustering}
     \label{tab:rating}
\end{table}
\FloatBarrier

\subsection{Explainability of the classification model using SHAP}
\label{subsection:explain}

To provide a comprehensive level of explainability, we use some of the charting tool provided by SHAP library: summary plot, dependence plots, and waterfall plot.
They give us a clear view of the model behavior both globally and locally.

\paragraph{SHAP summary plot}

\begin{figure}[h!]
     \centering
	\includegraphics[scale=0.6]{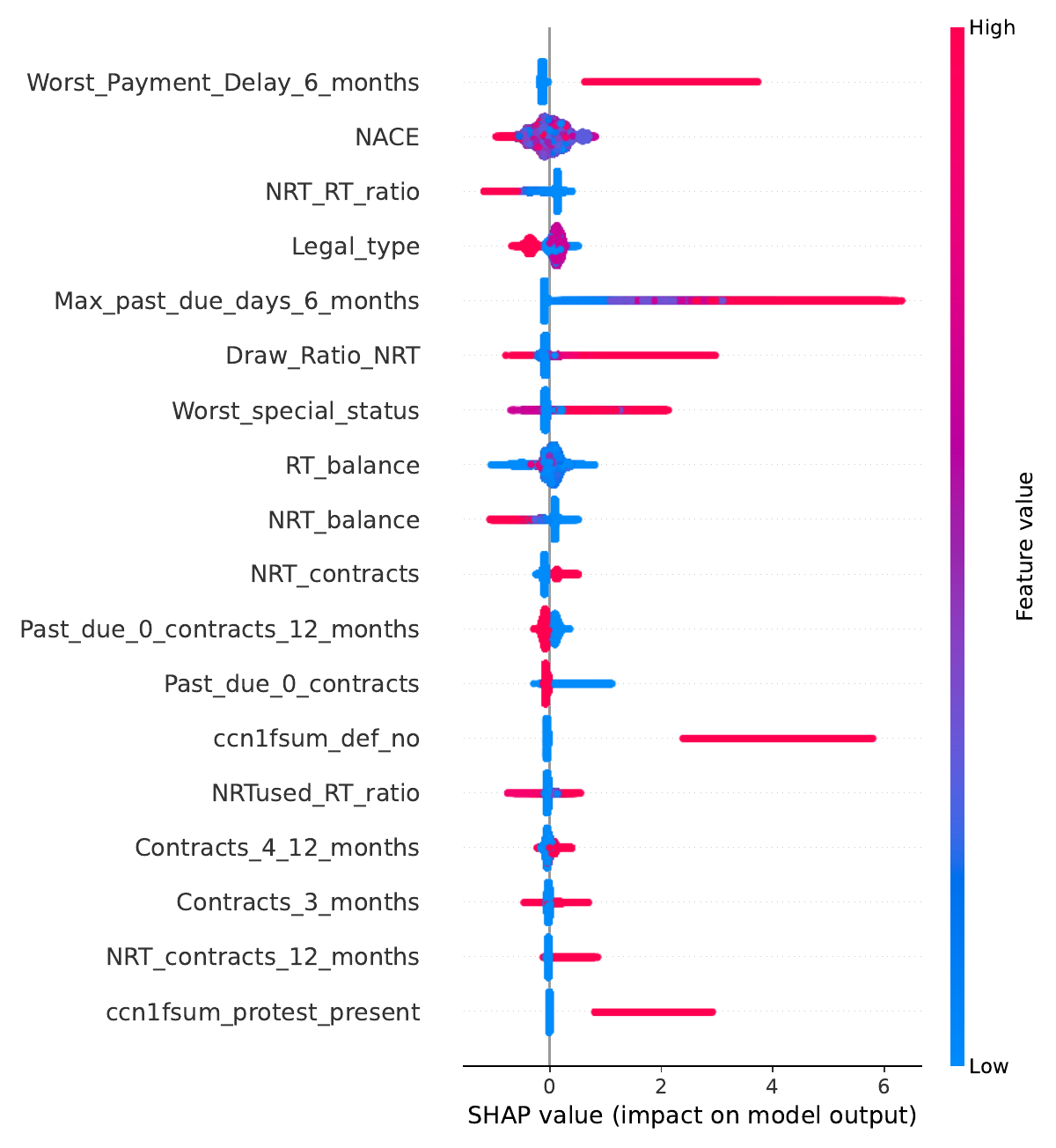}
     \caption{SHAP summary plot for Light-GBM classifier. }
     \label{fig:summary_plot}
\end{figure}

Figure \ref{fig:summary_plot} illustrates the overall impact of various features on the target. It is important to note that, in this application, a positive SHAP value indicates a higher probability of default, and vice versa.

As evident from the plot, a past due of one month significantly elevates the probability of subsequent default. Additionally, open credit lines correlate with a higher risk of default, as indicated by the positive SHAP values for the NRT contracts.

\FloatBarrier

\paragraph{SHAP dependence plots}
Shap dependency plots enhances information depicted on the SHAP summary plot (see \Cref{fig:summary_plot}) for single variables with respect to another related variable.

\begin{figure}[h!]
     \centering
     \subfloat[]{
     \includegraphics[scale=0.125]{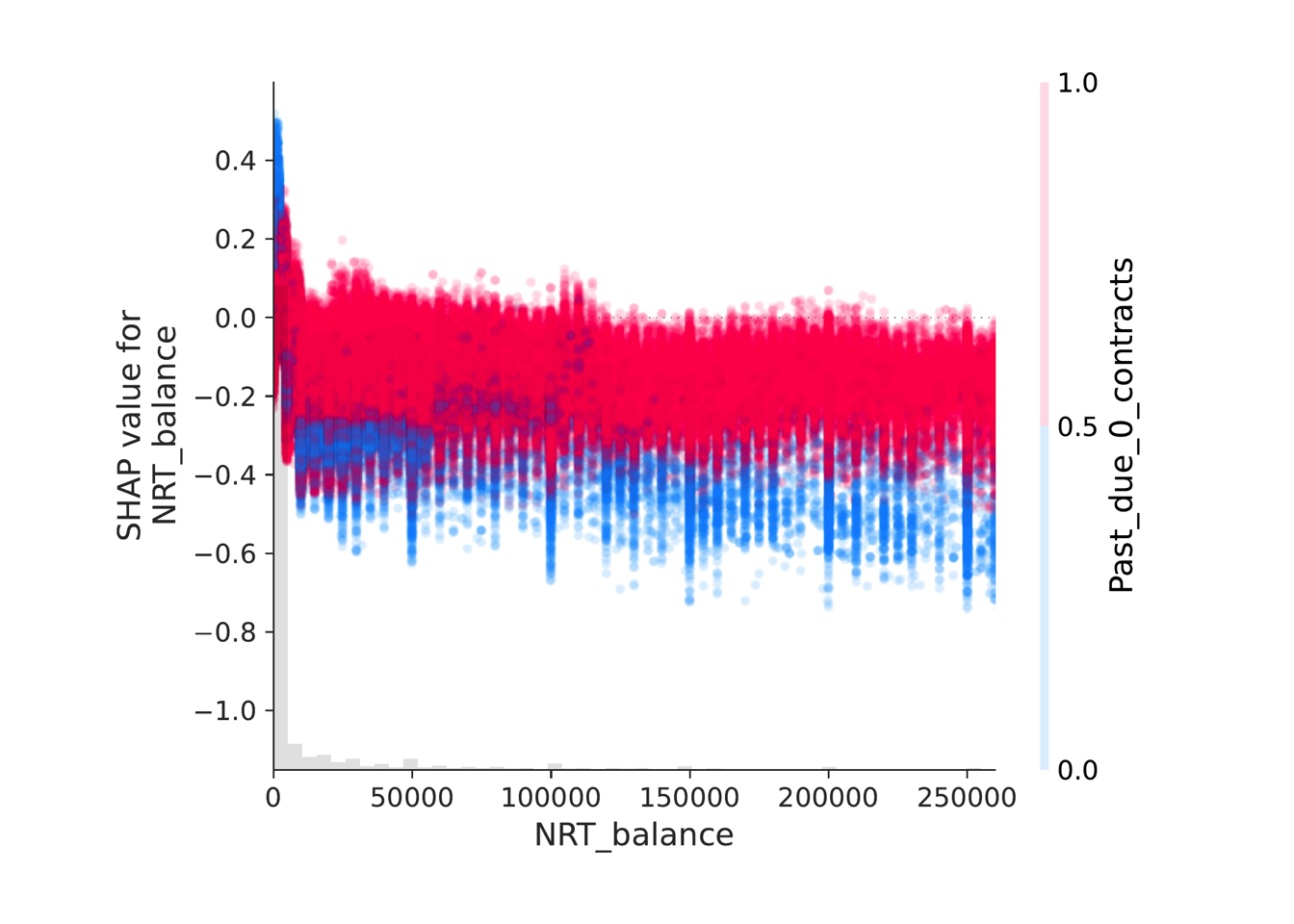}\label{fig:a}}
     \subfloat[]{\includegraphics[scale=0.125]{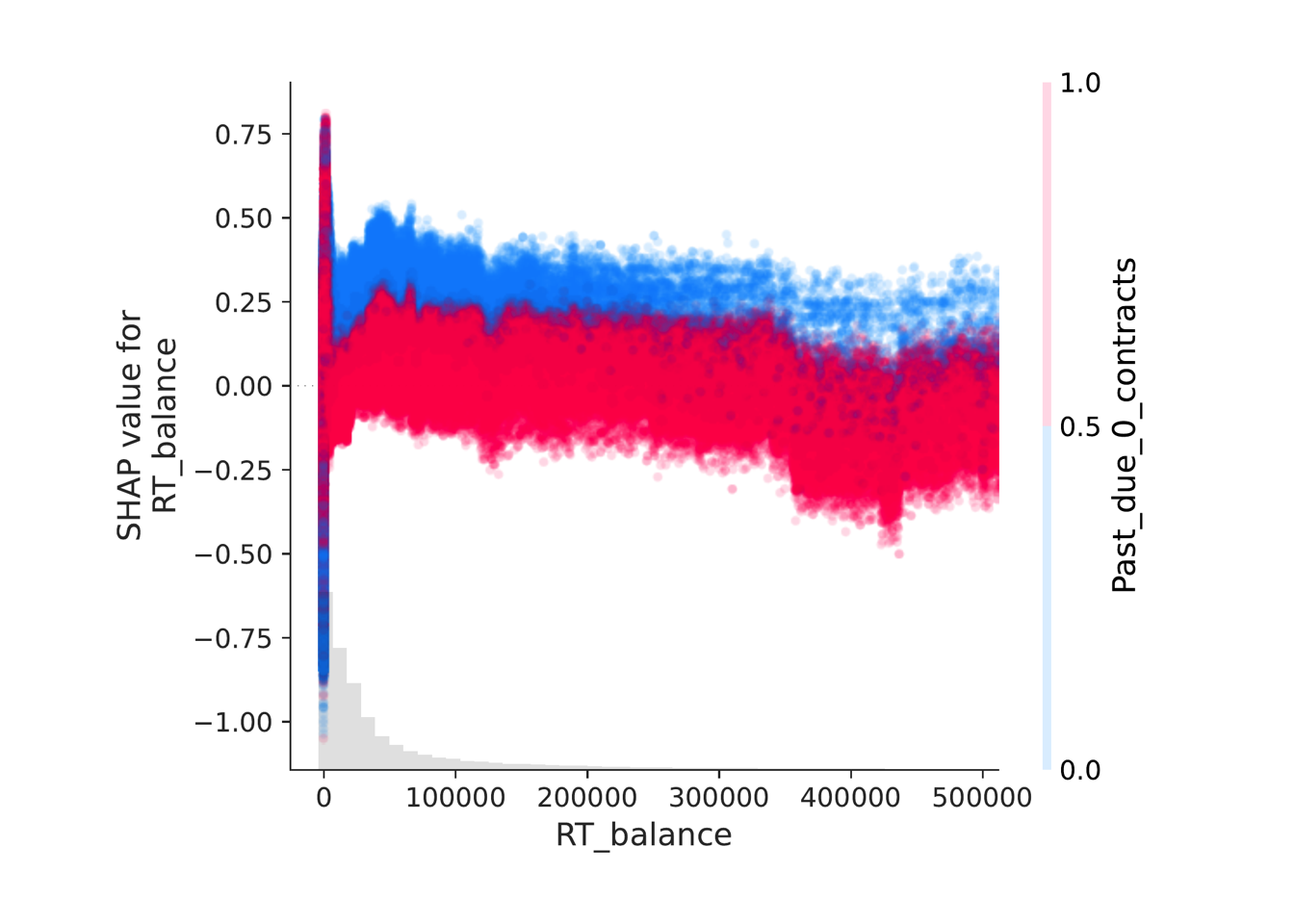}\label{fig:b}}\\
     \subfloat[]{\includegraphics[scale=0.125]{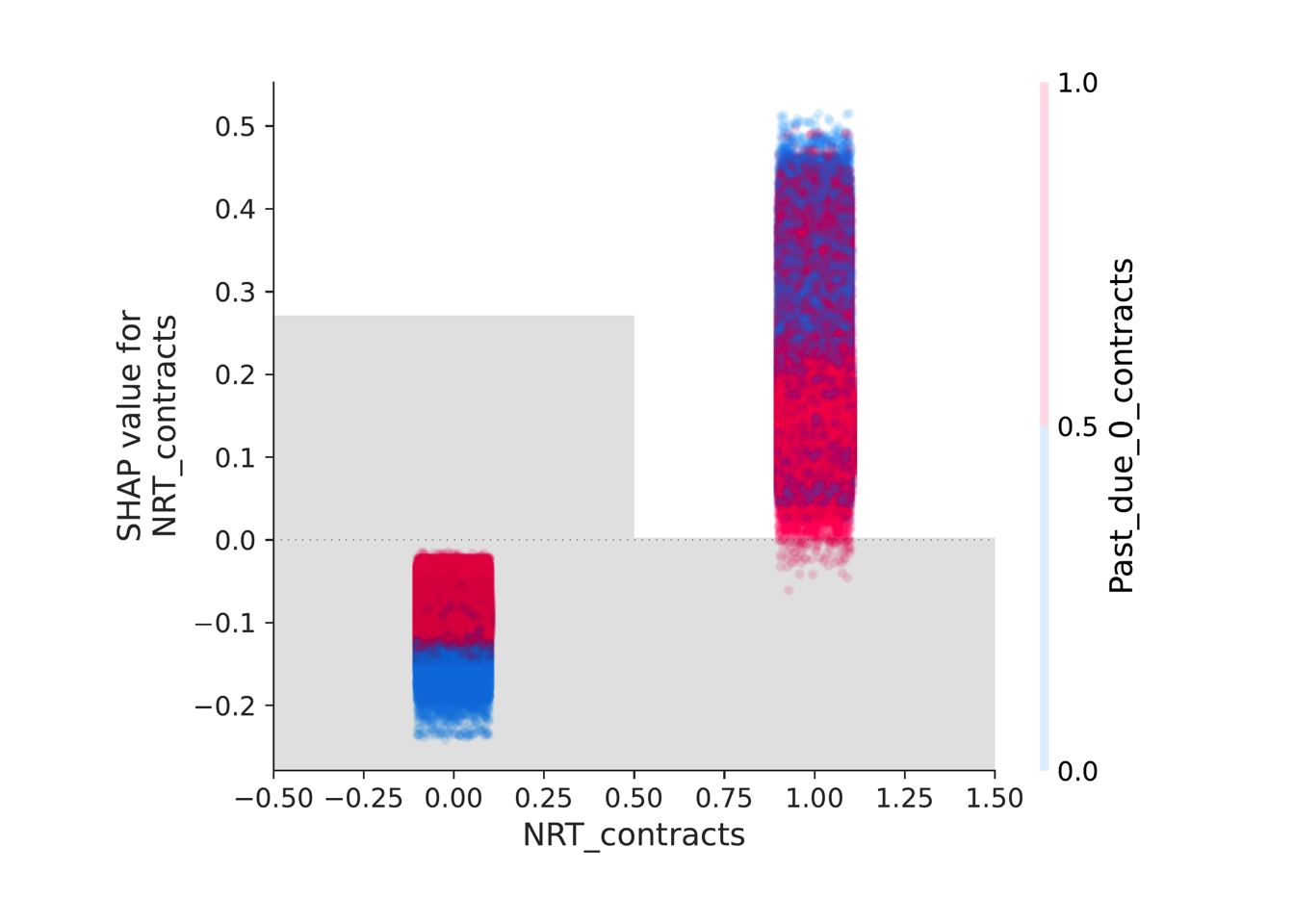}\label{fig:c}}
     \subfloat[]{\includegraphics[scale=0.125]{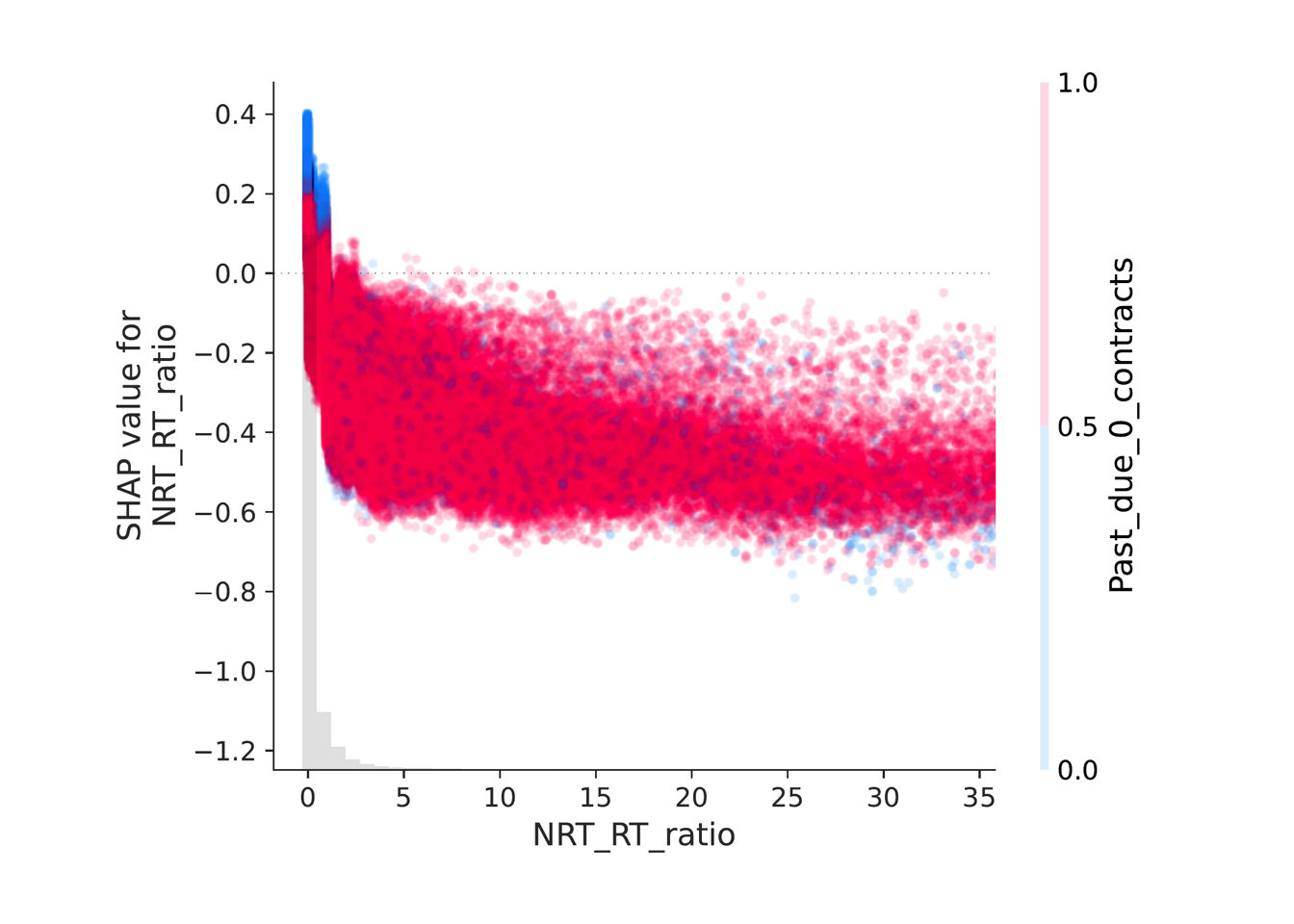}\label{fig:d}}\\
     \subfloat[]{\includegraphics[scale=0.125]{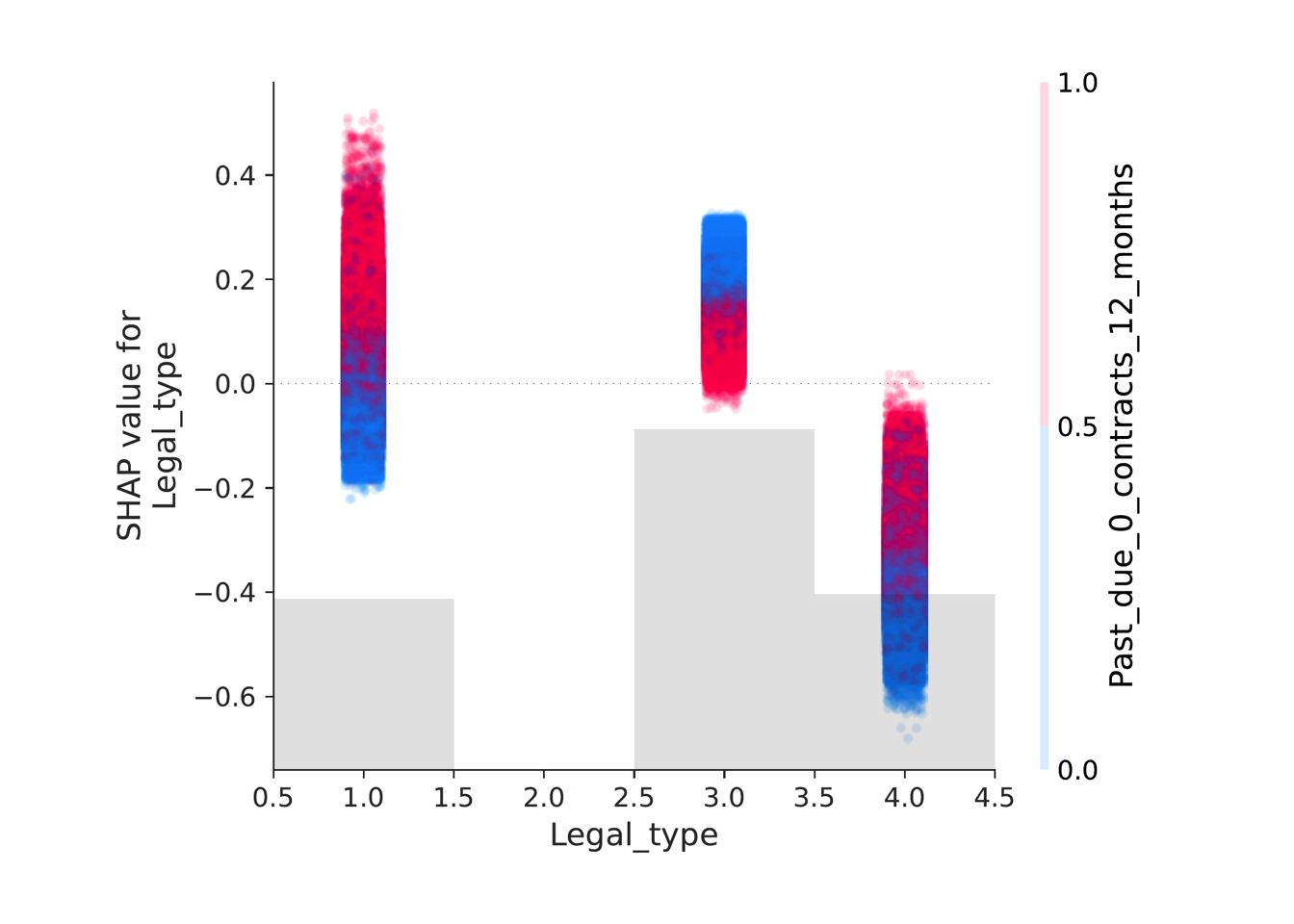}\label{fig:e}}
     \subfloat[]{\includegraphics[scale=0.125]{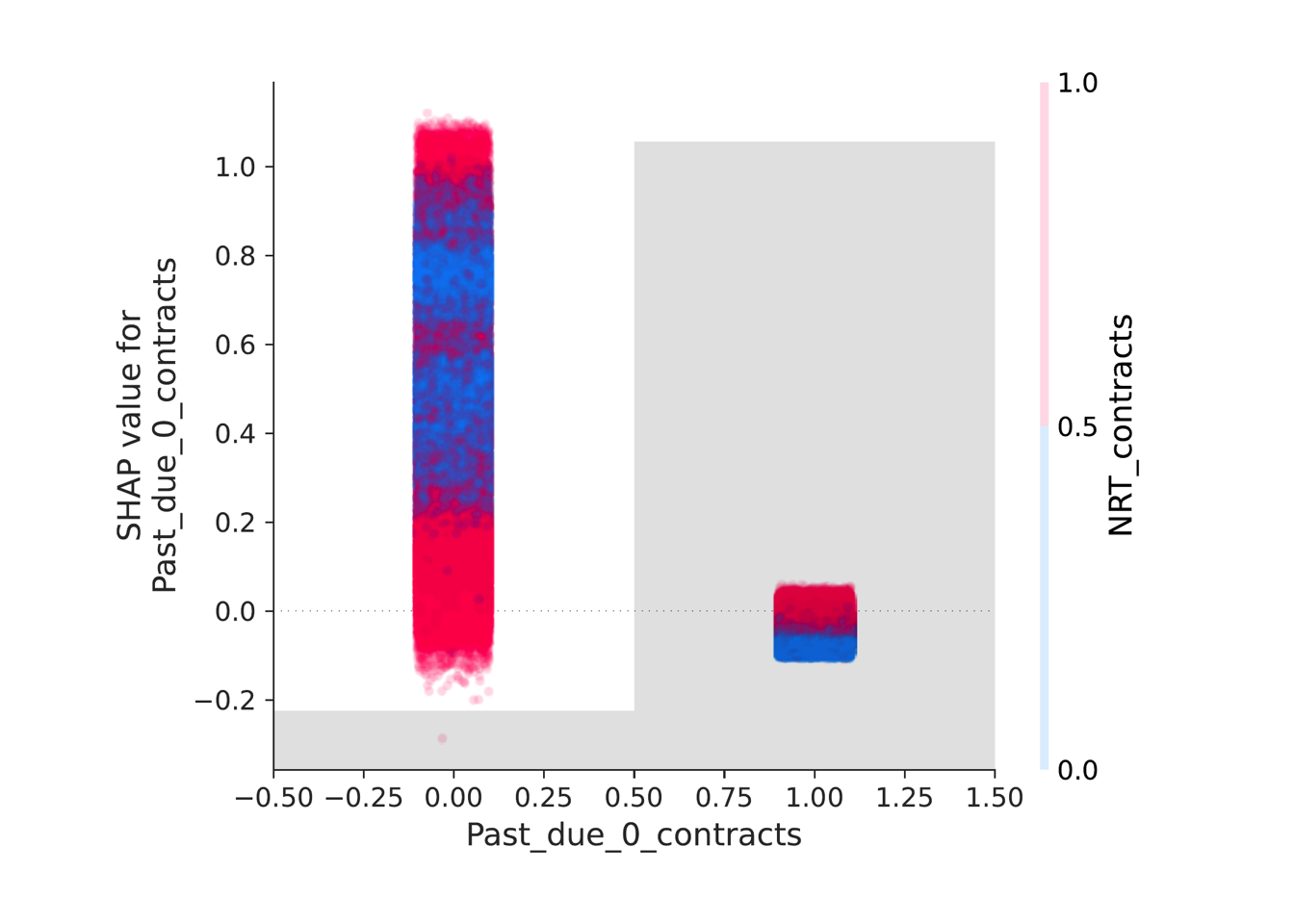}\label{fig:f}}\\
     \subfloat[]{\includegraphics[scale=0.125]{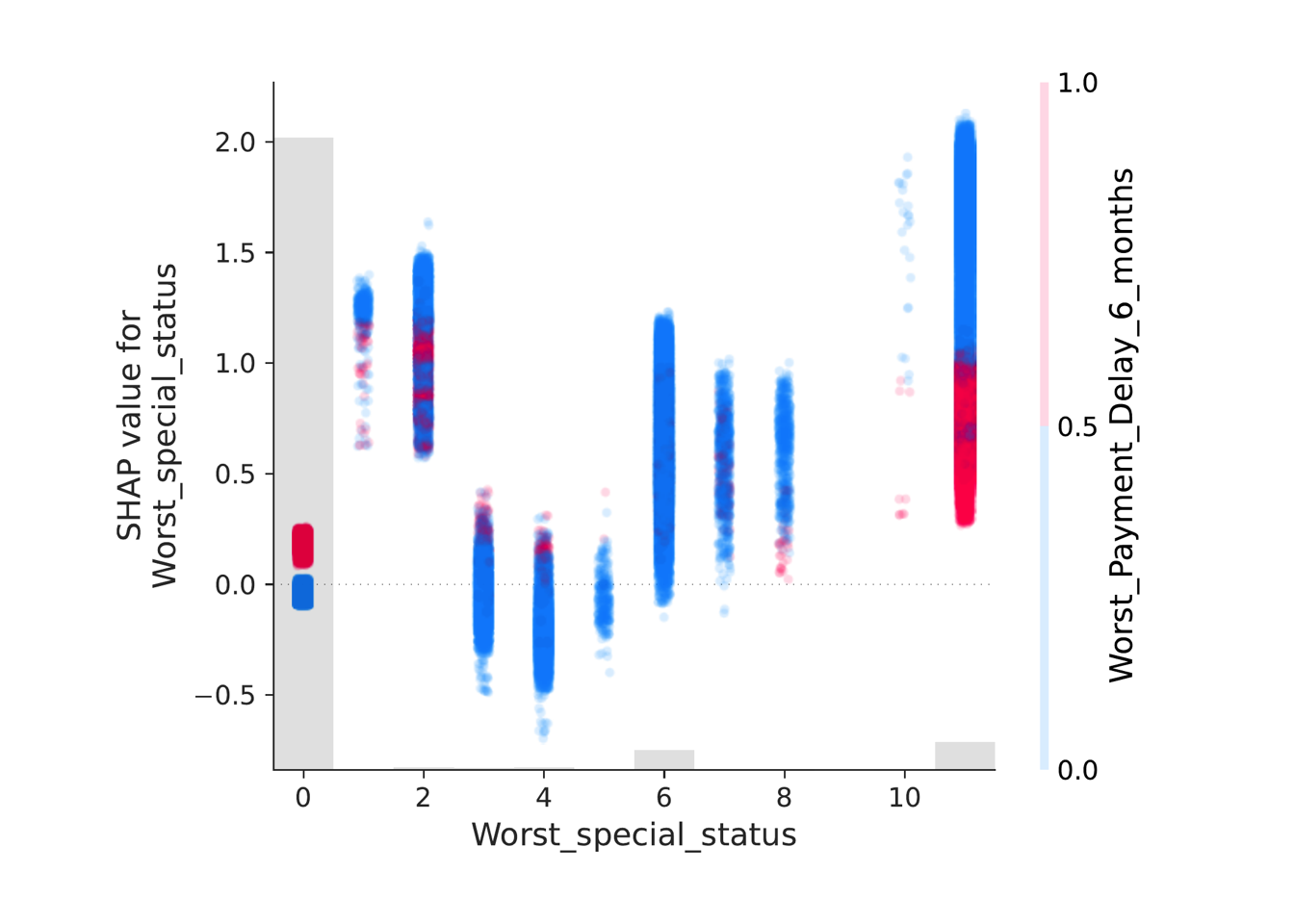}\label{fig:g}}
     \caption{SHAP dependency plots for NRT balance~\Cref{fig:a}, RT balance in~\Cref{fig:b}, NRT contracts in~\Cref{fig:c}, NRT RT ratio in~\Cref{fig:d}, Legal type in~\Cref{fig:e}, Past due 0 contracts in~\Cref{fig:f}, Worst special status in~\Cref{fig:g}. }
 \label{some-label}
\end{figure}

A notable and somewhat intricate behavior pertains to the legal type and the number of closed contracts with status 0. For limited companies (SC), this particular combination of features corresponds to lower probability of default. However, the situation is entirely opposite for sole proprietorships (DI) and partnership companies (SP).

One possible explanation for this phenomenon is that the absence of closed contracts in past due status can be linked to either of two situations: a) a lack of financial activity, or b) the presence of unfulfilled obligations.
The former is crucial for limited companies but not as significant for smaller entities like sole proprietorships or partnerships, whereas the latter impacts all types of entities.

\paragraph{An example of analysis: SHAP waterfall plot}

\begin{figure}[h!]
     \centering
	\includegraphics[scale=0.5]{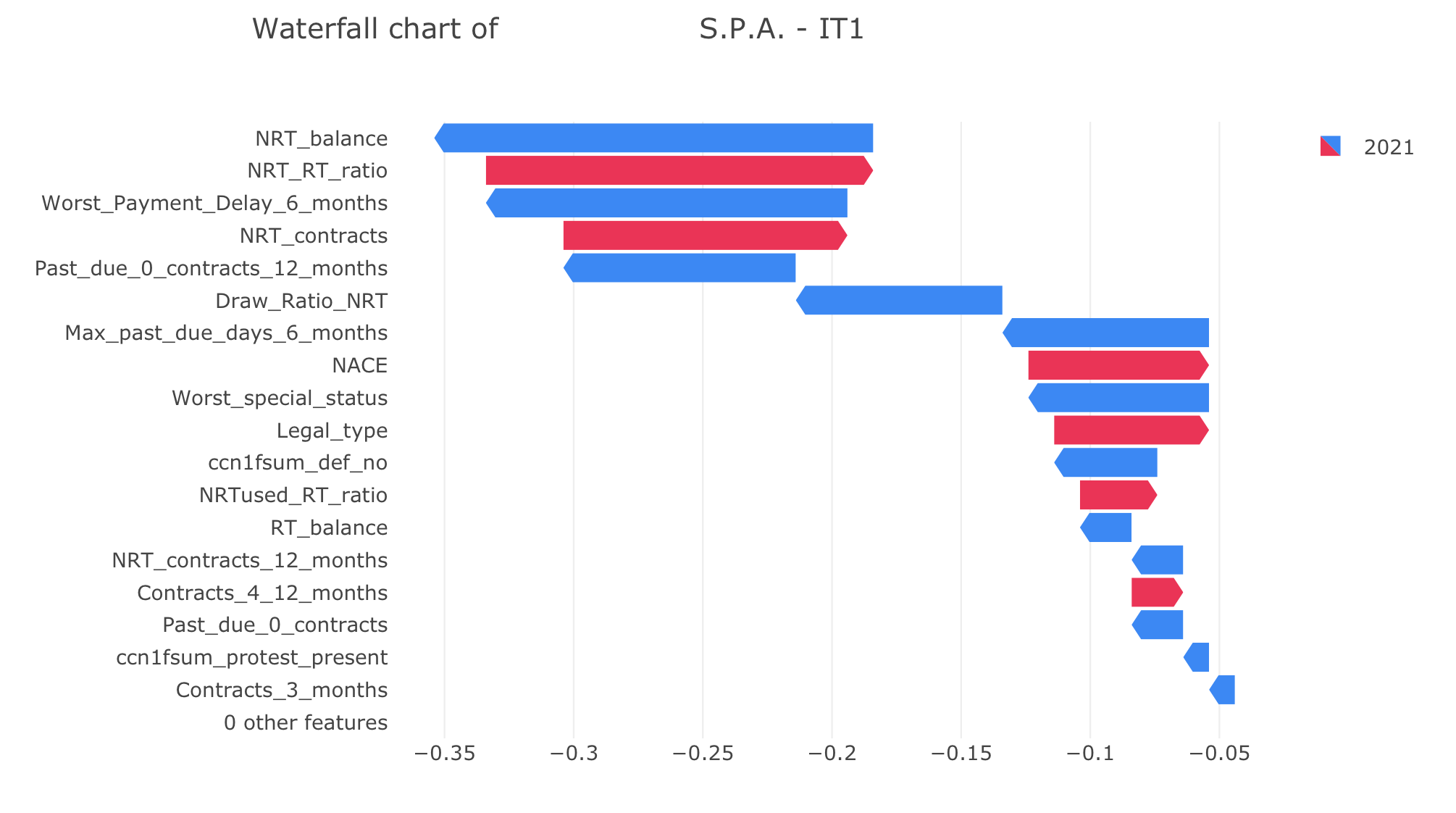}
     \caption{SHAP waterfall plot for an anonymous company}
     \label{fig: waterfall_plot}
\end{figure}

Figure \ref{fig: waterfall_plot} illustrates SHAP waterfall plot that contains individual features' effects on a single prediction.

It can be seen that the NRT balance significantly contributes to the decrease in the predicted PD.
At the same time both NRT to RT ratio and the indicator of NRT contracts increase the final PD.
This information can assist an analyst since it explains the marginal effects of indicators.

\FloatBarrier

\section{Application to the Central Credit Register}

The Central Credit Register (CR) is the official database of the Bank of Italy, recording information about all legal entities borrowing money from financial institutions in Italy.
Regulatory requirements mandate these institutions to contribute information about open credit lines to the Bank of Italy on a monthly basis.
Subsequently, the Bank of Italy consistently updates all financial institutions with information on the credit activity of their debtors.

In contrast, Experian, functioning as a private credit bureau, collects data exclusively from cooperating banks. While Experian reports more detailed information, both the CR database and Experian are expected to be grounded in the same underlying information, despite differences in their data models.

Although the features presented in the two datasets are not exactly identical due to not all CR data being contributed to Experian and the less granular nature of CR updates compared to Experian, our objective is to align the Central Credit Register data with the Experian format, considering all discussed limitations.
Consequently, we aim to leverage the CR data by utilizing mapped features within our model trained on the Experian data.
The successful implementation of data models' alignment finalizes the transfer learning procedure.

\subsection{CR Data}

The Bank of Italy employs two mechanisms for sharing information about existing contracts and balances with financial entities: monthly updates and preliminary information communication.
Through monthly updates, banks receive regular information about their debtors. In contrast, preliminary information communication is specifically triggered by a new loan request, aiming to validate potential clients by sharing relevant CR data.
While both mechanisms communicate similar data, their goals differ. In the first case, the objective is risk control or debtor supervision, whereas in the second case, it involves the evaluation of credit risk and credit pricing for subsequent lending (\quota{fase di valutazione e dell'erogazione di credito}).

CR data consist of 18 credit surveys \cite{cr1991}, grouped into 5 categories: cash loans (\quota{crediti per cassa}), unsecured loans (\quota{crediti di firma}), received collaterals (\quota{garanzie ricevute}), financial derivatives (\quota{derivati finanziari}), and information section (\quota{sezione informativa}). Since the features discussed in Section 1.3.3 are related to mortgages/installments and open contracts, our focus is on maturity risks and revocable credit risks, both of which are part of cash loans. For more information about the CR and communications of the Bank of Italy, readers are referred to Circular 139 \cite{cr1991}.

Contracts are summarized by 8 variables, including the original duration of a contract, remaining duration, type of contract, state of a contract, etc. Each of these variables is qualitative, with several categories. For each possible combination, the CR reports the total sum of balances, amounts used, and past due amounts. For instance, the original duration has 3 levels: a) less than a year, b) from 1 to 5 years, and c) more than 5 years. Remaining duration has only two levels: more than a year and less than a year. Note that, when working with open credit lines (revocable credit risks), original and remaining durations are not available and therefore not provided. Finally, the state of a contract indicates 90-days delinquencies and 180-days delinquencies.

\subsection{CR features mapping}
Based on the CR information discussed above, we identified two groups of features that should be treated differently.
The first one contains variables that can be directly transformed to the experian format.
Consequently, elements of the other one can be mapped only to a proxy feature, that has similar meaning to the experian counterpart.

Here we provide a set first group features with the mapping description:
\begin{itemize}
     \item \quota{NRT contracts} -  Total balance of revocable credit risks is more than 0
     \item \quota{NRT balance} - Total balance of revocable credit risks
     \item \quota{NRT used} - Total used balance of revocable credit risks
     \item \quota{NRT past due credit balance} - Total overdue of revocable credit risks
     \item \quota{Closed NRT} - At least one decrease of balance of revocable credit risks during last 12 months
     \item \quota{RT past due}(\quota{Worst\_Payment\_Delay\_6\_months}):

     \item NPL present  - Presense of at least 1 NPL (Phenomenon \quota{000551000})
     \item \quota{Protesti} (\quota{protest\_present}) - this information is taken from another database
     \item \quota{Legal type} - this information is taken from internal database
     \item \quota{RT balance} - Total balance of maturity risks
     \item \quota{NRT RT ratio} - ratio between NRT balance and RT balance
     \item \quota{NRTused RT ratio} - ratio between  used NRT balance and RT balance
     \item \quota{Draw Ratio NRT} - ratio between used NRT balance and NRT balance
\end{itemize}

To map past-due we begin with acknowledging that presence of small amounts in past due
may be motivated by the processing issues and does not necessarily correspond to a higher risk of default.
Indeed, the Bank of Italy adopts absolute and relative materiality thresholds for the proper definition of default.
These values are 500 euro and 1\% of total balances.
If an absolute or relative balance of past due is less than these values a company remains 'in bonis'.
Hence, we propose a two-step approach for past due detection based on these thresholds, for the sake of precision and concordance with the regulation.

The initial step involves calculating the ratios of $past\_due\_0$, $past\_due\_30$, $past\_due\_90$, $past\_due\_180$. The subsequent step entails comparing these calculated values against predefined thresholds. It is only after this comparison that we can accurately specify a company's past due status.

We calculate past due status as follows:
\begin{itemize}
     \item $past\_due\_180\_balance$ is a sum of the balances of credit lines with statuses (827, 831, 125, 129, 133, 137)
     \item $past\_due\_90\_balance$ is a sum of the balances of credit lines with statuses (826, 830, 124, 128, 132, 136)
     \item $past\_due\_30\_balance$ is a sum of past due values of all other credit lines
     \item $past\_due\_0\_balance$ is a sum of the balances of all other credit lines minus $past\_due\_30\_balance$
\end{itemize}

Note that the past dues are calculated separately for NRT and RT products in order to mimic experian structure.

Following features represent the second group:
\begin{itemize}
     \item \quota{Special status} - only \quota{dispute} and \quota{insolvency} are indicated within the CR data
     \item Closed past due 0 (\quota{past\_due\_0\_contracts\_12\_months} $>=1$) - Note that the communications of the Bank of Italy contain aggregated imports and not values for each single credit line. Hence, it is only possible to verify that overall past due is 0.
     To indicate closure without past due during last 12 months, we propose to use a proxy feature. We check  a) decrease of balance and b) that past due is equal to 0 during both the recent one and the previous months.

     \item \quota{Contracts 3 months} - at least one increase of balance of maturity risks during last 3 months, without subsequent closure
     \item \quota{Contracts 4-12 months} - at least one increase of balance of maturity risks between 4 to 12 months ago, without subsequent closure
     \item \quota{Past due 0} (\quota{Past\_due\_0\_contracts} $>= 1$) - past due amount of at least one group of variables of classification is equal to 0
\end{itemize}

\subsection{Mapping validation}

Given the degree of uncertainty associated with the proposed mapping, a thorough validation is conducted through a dual-strategy approach, each with its distinct focus and methodology.

The initial step involves an in-depth examination of the connection between Experian and CR mapped features. Subsequently, we assess the consistency between the predictions derived from these two datasets.

The second step is a temporal validation through backtesting. Here, historical data is utilized to evaluate the model's ability to predict default. This stage not only tests the model's predictive accuracy but also reinforces the effectiveness of the mapping in operational settings.

\subsubsection{Comparison with the Experian data}
Conducting a rigorous validation of the proposed mapping, it is essential to acknowledge that the Experian dataset is constructed from a subset of data available within the Central Credit Register (CR) \cite{cr1991}.
Given that CR aggregates comprehensive data across all reporting entities,
it should always contain a broader set of credit lines and associated financial indicators compared to Experian's more narrowly sourced data.

Here validation strategy is based on a carefully created dataset of 180 companies that are presented in both CR and Experian datasets,
with a deliberate focus to match reference periods.
This process ensures that credit lines, balances, and binary indicators from Experian are consistently mirrored or outnumbered by those in CR,
adhering to the premise of this part.

To address potential variances arising from the distinct data models of CR and Experian a robust statistical framework is essential.
We start with questioning what if our premise is not correct.
Thus, for numeric features such as RT or NRT balances,
the alternative hypothesis  is that Experian values would not exceed those of CR.
Our analyses aimed at rejecting the null hypotheses through paired sample tests, affirming proper relationship.
For binary variables, a similar approach was adopted, with expectations set for CR to display an equal or greater frequency of such events.
Due to non-normality, the first choice for numeric features testing is the non-parametric Wilcoxon signed-rank test \cite{conover1999practical}.
We then verified the interdependence of the values with Spearman's $\rho$, confirming the expected correlation.
In the binary case we utilized McNemar's \cite{mcnemar1947note} test to assess the differences in outcomes for the paired data.

Considering the complexity of conducting multiple statistical tests on linked data,
we employed the False Discovery Rate (FDR) control method \cite{benjamini1995controlling}.
In particular, the Benjamini-Yekutieli (BY) \cite{benjamini2001control} procedure was chosen due to its capability to control the FDR under any dependency structure.
This conservative approach to FDR control adds a layer of strict control to validation process, ensuring that our findings are robust and reliable.

\begin{table}[h!]
     \centering
     \begin{tabular}{@{}cccc@{}}
     Feature & p-value & adjusted p-value & Result \\
     NRT contracts 12 months &  $2.5 \times 10^{-22}$ & $2 \times 10^{-21}$  & Reject Null \\
     Contracts 3 months & $2.4 \times 10^{-11}$ & $1.9 \times 10^{-10}$  & Reject Null \\
     Worst Payment Delay 6 months & $3 \times 10^{-6}$ & $2.4 \times 10^{-5}$  & Reject Null \\
     Past due 0 contracts & $1.8 \times 10^{-7}$ & $1.4 \times 10^{-6}$  & Reject Null \\
     NRT contracts & $7 \times 10^{-16}$ & $5.6 \times 10^{-15}$ & Reject Null \\
     RT balance & $2.6 \times 10^{-27}$ & $2.1 \times 10^{-26}$ & Reject Null \\
     NRT balance & $2.5 \times 10^{-5}$ & $2 \times 10^{-4}$ & Reject Null \\
     Max past due days 6 months & $5.5 \times 10^{-14}$ & $4.4 \times 10^{-13}$ & Reject Null \\
     \caption{Hypothesis Testing Results with FDR Correction (BY)}
     \label{tab:hypothesis_testing}
     \end{tabular}
\end{table}

Table \ref{tab:hypothesis_testing} reports the p-values of described statistical tests and the adjusted p-values based on the BY procedure.
Each adjusted p-value is less than 0.0001, hence  all null hypotheses are rejected.
This outcome affirms that Experian data are a subset of CR.

Finally, we check the agreement between predictions based on the CR and Experian datasets with the Kendall's tau.
The statistic has value 0.25 and is significant, indicating a partial but meaningful correspondence between predictions.
This level of concordance, though not high, aligns with our expectations and confirms a noticeable ordinal association.

\subsubsection{Validation on historical data}
In a secondary evaluation of the mapping, we examine the model's capacity to predict instances of past due 90 days or default within a one-year timeframe.
This analysis involves historical data for companies with known statuses, creating a timeframe that aligns with the backtesting procedure.
Two comparable groups of companies are identified, mirroring those used during training.

The first group consists of companies with a minimum of two years of financial records, exhibiting no past due 90 days or bankruptcy in the preceding year.
The second group comprises companies with at least 90 days of delinquencies or bankruptcy within the past year.
Notably, predictions are made using records from the penultimate year, ensuring consistency between the training and validation stages.

Upon querying the Central Credit Register, we identified 1605 companies labeled as ``in bonis'' and 367 companies categorized as 'past due 90 or worse'.
The resulting 18.6\% validation default rate is approximately five times higher than the 3.5\% training default rate. This discrepancy can be attributed to several factors: the Central Credit Register encompassing all available information about existing and active credit lines, Experian containing data contributed only by banks with agreements, the exclusion of companies with less than 2 years of records, and illimity Bank's focus on a niche of companies with a higher risk of past dues.

Table \ref{tab:validation_confusion_matrix} presents the normalized confusion matrix for the LightGBM classifier, with bold indicating absolute numbers.
The matrix closely mirrors the training confusion matrix in Table \ref{tab:class_confusion_matrix}. During the validation phase, a slightly higher rate of true positives (0.3 compared to 0.33 in training) and a small increase in false negatives (0.11 versus 0.09 during training) are observed.
Classification metrics are detailed in Table \ref{tab:backtesting_metrics_}.

\begin{table}[h!]
     \centering
     \begin{tabular}{lcc}
          & Predicted Negative (in bonis) & Predicted Positive (defaulted)   \\
          Actual Negative (in bonis) &  0.70 (\textbf{1116})& 0.30 (\textbf{489})  \\
          Actual Positive (defaulted)  & 0.11  (\textbf{40}) & 0.89 (\textbf{327}) \\
     \end{tabular}
     \caption{Normalized confusion matrix for the classifier prediction, bold indicates numbers of companies}
     \label{tab:validation_confusion_matrix}
\end{table}

\begin{table}[h!]
     \centering
     \begin{tabular}{|c|c|}
     \hline
     \textbf{Description} & \textbf{Value} \\ \hline
     AUC & 0.903 \\ \hline
     Recall & 0.891 \\ \hline
     Specificity & 0.695 \\ \hline
     $F_{\beta}$ Measure & 0.795 \\ \hline
     Average Precision & 0.757 \\ \hline
     \end{tabular}
     \caption{Backtesting metrics}
     \label{tab:backtesting_metrics_}
\end{table}

A comparative analysis of these metrics alongside the training phase metrics (see Table \ref{tab:metrics}) indicates that the classifier's performance during validation closely aligns with its performance in training.

\section*{Conclusion}
\addcontentsline{toc}{section}{Conclusion}
\label{section:conclusion}

In this paper, we present a stack of models designed to estimate the 1-year credit rating of corporate entities by utilizing both the regulatory default definition and behavioral data from credit bureaus.
The architecture builds upon the framework established in \cite{provenzano2020machine}, with notable enhancements such as a more effective feature selection process and improved explainability.

Our model demonstrates competitive performance compared to existing solutions, achieving results that are in line with industry standards.

A distinctive contribution of this paper is the novel approach to data mapping from the Central Credit Register (CR), the database of the Bank of Italy, to the Experian credit bureau.
We meticulously outline the mapping approach between these databases, enabling us to harness both sources for inference.
Then, we implement a robust dual strategy validation procedure to ensure the reliability and applicability of our proposed mapping within our predictive model.

The initial validation step confirms that Experian data, despite having a narrower scope, constitutes a reliable subset of the CR dataset, establishing a partial predictive concordance.
The second step involves backtesting with historical data, affirming the model's accuracy in predicting financial defaults and highlighting its robustness across diverse credit histories.
The consistent performance of our model underscores its real-world applicability.

Together, these validation steps not only showcase the effectiveness of our mapping but also validate the appropriateness of the transfer learning procedure presented in our study.
This guarantees the reliability and versatility of our model in leveraging different data sources, enabling a more comprehensive credit risk assessments.

\newpage

\section*{Acknowledgements}
We thank Lorenzo Giada for insightful feedbacks, Luca Massaron and Federico Zaniolo for support with the CR analysis.

%%%%%%%%%%%%%%%%%%%%%%%%%BIBLIOGRAPHY%%%%%%%%%%%%%%%%%%%%%%%%%%%%%%%%%%%%%%%%%%%
\bibliographystyle{plain}
\bibliography{behavioral_paper}

%%%%%%%%%%%%%%%%%%%%%%%%%APPENDIX%%%%%%%%%%%%%%%%%%%%%%%%%%%%%
\newpage
\newpage

\appendix

\end{document}